\documentclass[showpacs,twocolumn,amsmath,amssymb,pre,aps,floatfix]{revtex4}
\usepackage{epsfig}
\usepackage{graphicx}
\usepackage{dcolumn}
\newcommand{\concrate} [2] {\ensuremath {\frac{d [n_{\text {\tiny #1}}^{\text {\tiny #2}} ]}{dt}}}
\def \concacid {\ensuremath {[n_{\text {\tiny COOH}}]}}
\def \concohptmg {\ensuremath {[n_{\text {\tiny OH}}^{\text {\tiny PTMG}}]}}
\newcommand{\conctmp} [1] {\ensuremath {[n_{\text {\tiny TMP}}^{\text {\tiny #1}}]}}
\def \molptmg {\ensuremath {n_{\text {\tiny PTMG}}}}
\def \moltmp {\ensuremath {n_{\text {\tiny TMP}}}}
\def \molad {\ensuremath {n_{\text {\tiny AD}}}}
\begin{document}
\title{Dynamic scaling in entangled mean-field gelation polymers}
\author{Chinmay Das$^{1,2}$, Daniel J.\ Read$^{1}$, Mark A.\ Kelmanson$^{1}$ 
and Tom C.~B.\ McLeish$^{2}$}
\affiliation{$^1 \;$ Department of Applied Mathematics, University of Leeds,
Leeds, LS2 9JT, U.K. \\
$^2 \;$ Department of Physics and Astronomy, University of Leeds,
Leeds, LS2 9JT, U.K.}
\date{\today}
\pacs{82.70.Gg, 
83.10.Kn, 
02.70.-c 
}

\begin{abstract}
We present a simple reaction kinetics model to describe the
polymer synthesis used by Lusignan {\em et~al.\ }\protect{\cite{lusignan:99}}
to produce randomly branched polymers in the vulcanization class.
Numerical solution of the rate equations gives probabilities for
different connections in the final product, which we use to 
generate a numerical ensemble of representative molecules.
All structural quantities probed in ref.~\protect{\cite{lusignan:99}} are in
quantitative agreement with our results for the entire range of
molecular weights considered. However, with detailed topological
information available in our calculations, our estimate of
the `rheologically relevant' linear segment length is smaller than that
estimated in ref~\protect{\cite{lusignan:99}}.
We use a numerical method \protect{\cite{das:06}} based on tube
model of polymer melts to calculate the rheological
properties of such molecules.  Results are in good agreement with
experiment, except that in the case of the
 largest molecular weight samples
our estimate of the zero-shear viscosity is significantly lower than the
experimental findings. Using acid concentration as an indicator for closeness
to the gelation transition, we show that the high-molecular-weight polymers
considered are at the limit of mean-field behavior - which possibly
is the reason for this disagreement. For a truly mean-field gelation
class of model polymers, we numerically calculate the rheological properties
for a range of segment lengths. Our calculations show that the tube
theory with dynamical dilation predicts that, very close to the
gelation limit, contribution to viscosity for this class of
polymers is dominated by the contribution from constraint-release 
Rouse motion and the
final viscosity exponent approaches Rouse-like value.
\end{abstract}

\maketitle
\section{Introduction}
Polycondensation reactions that generate branched polymers 
lead to progressively larger molecules as a function of the
conversion. The same thing happens during chemical cross linking 
(vulcanization).
At a critical extent of the reaction or density of bonds $p_c$, the 
size of the largest molecule spans the system and this is termed
as the gel point \cite{stauffer:92,degennes:79,rubinstein:colby}. 
Close to the gel point, static properties of the
system exhibit a scaling form. Defining $\epsilon = \frac{|p - p_c|}{p_c}$,
the number fraction $\Phi(M)$ of the molar mass $M$ falls off as a power law,
\begin{equation}
\Phi(M) \sim M^{-\tau} f(M/M_{char}),
\label{eq:massscaling}
\end{equation}
where $f$ is a cut-off function and the characteristic mass $M_{char}$
diverges as
\begin{equation}
M_{char} \sim \epsilon^{-1/\sigma}.
\end{equation}
Different moments of the molar mass distribution also diverge as 
the gel point is approached. In particular, the weight-averaged molar
mass $M_W$ diverges as
\begin{equation}
M_W \sim \epsilon^{-\gamma}.
\end{equation}
The static exponents $\tau$, $\sigma$ and $\gamma$ depend on the
universality class for a given system. When the molecules in the
melt overlap strongly (as in vulcanization of long linear molecules),
the exponents belong to  the mean-field universality class and are described by
Flory-Stockmayer theory \cite{stockmayer:43,flory:53}. 
In this case, the exponents can be
calculated analytically with $\tau = 5/2$, $\sigma = 1/2$ and
$\gamma = 1$. Polymerization of small multifunctional groups
lead to the critical percolation gelation class. Though exponents
for this class cannot be calculated analytically, good
estimates for the exponents are known from simulations \cite{alder:90}.

Close to the gel point, under some circumstances, the rheological 
properties also
obey scaling forms \cite{valles:79,stauffer:83,durand:87,winter:87,martin:89,nicol:01,gasilova:02}.
The shear relaxation modulus $G(t)$ is
a power law in time ($t$) and the complex viscosity $\eta^*(\omega)$ is a power law in 
frequency ($\omega$):
\begin{eqnarray}
G(t) &\sim& t^{-u} \;\;\;\; {\rm and} \nonumber \\
\eta^*(\omega) &\sim&  \omega^{u-1}.
\end{eqnarray}
The zero-shear viscosity $\eta$ diverges with exponent $s$ and the recoverable 
compliance $J_e^0$ diverges with exponent $t$:
\begin{eqnarray}
\eta &\sim& \epsilon^{-s} \; \; \;\; {\rm and} \nonumber \\
J_e^0 & \sim &  \epsilon^{-t}.
\end{eqnarray}

The dynamic exponents are not derivable from the static ones 
without further assumptions \cite{cates:85} 
and for entangled polymers
the effective exponents (since the relaxation is only
approximately a power law \cite{rubinstein:90})
depend on the length of the linear segments between branch-points.
By considering simple rules for relaxation as a function of the  
{\em seniority} variables of the segments in a branched material,
the variation of the dynamic exponents
as a function of segment length has been estimated \cite{rubinstein:90}. 
A detailed calculation which takes care of
the molecular topology without such approximations is missing. 
In a recent publication \cite{das:06},
we used a numerical method to calculate the relaxation of 
arbitrarily branched material
within the broad framework of tube theory \cite{doi:book} and its extensions 
to handle constraint release \cite{marrucci:85,ball:89,colby:90}
and constraint-release Rouse motion \cite{viovy:91,milner:98}.
Relaxation of branch-on-branch architectures 
were included in a manner which respects the polydispersity both in
length and in topology. 
In this paper we attempt to use such a numerical scheme to
calculate the rheological relaxation function and 
the dynamic exponents close to the gel point.

The rest of the paper is organized as follows.
In a recent paper, Lusignan {\em et~al.\ }\cite{lusignan:99} reported 
synthesis and 
characterization of a series of randomly branched polyesters which are in the 
mean-field gelation class.  After a brief description of their reaction scheme
(sec.~\ref{sec:kinetic}), we use a 
simple kinetic model to determine the various probabilities for the connectivity in the 
final product. Using such probabilities we generate representative molecules and characterize 
the static structural properties and compare them with the experimental 
findings (sec.~\ref{sec:ptmg:static}). We provide a brief qualitative
description of the numerical method of \cite{das:06} to calculate
the rheological properties of branched entangled polymers in 
sec.~\ref{subsec:comprheo}.
In sec.~\ref{subsec:ptmg:expon} we calculate the linear rheological response
of the molecular ensembles and compare with the experimental results. 
The average inter-branch-point segment length in these polymers 
depends on the extent of esterification and the estimate of
this length is subject to errors. 
In sec.~\ref{sec:gelation} we consider a simplified ensemble of molecules 
which have predetermined average 
segment length and investigate the segment length dependence of the dynamical exponents. We conclude the paper by recapitulating the main findings 
of this study and stressing the questions this work raises on our 
understanding of the relaxation of highly branched polymers at the
longest timescales.

\section{Kinetic modeling for branched PTMG polymers \label{sec:kinetic}}
\begin{figure}[htbp]
\centerline{ \includegraphics[width=6cm,clip=true]{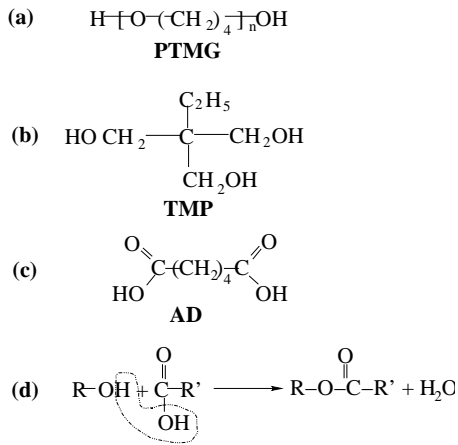}}
\caption{(a)-(c) Molecular structure of the reactants in polycondensation
reaction considered in \cite{lusignan:99} (d) Schematic description
of the esterification involved in the synthesis.} 
\label{fig:reactionscheme}
\end{figure}

In a recent study, Lusignan {\em et~al.\ }\cite{lusignan:99} considered a
polycondensation reaction of a polytetramethylene glycol (PTMG) oligomer with 
number-averaged molar mass
$M_N = 2900$~g/mol, trimethylolpropane (TMP) and adipic acid (AD). 
The two OH groups
at the ends of PTMG and three OH groups at the ends of TMP molecules (see figure
\ref{fig:reactionscheme}) react with the two acid groups of AD. 
Thus AD works as a bridging molecule connecting the TMP and PTMG molecules. 
The trifunctionality of TMP
molecules leads to branching. FASCAT 4100 (monobutyl tin oxide) was used as
catalyst which becomes incorporated in the final product. In our simplified
description, we assume that the catalyst simply increases the reaction
rate without changing the final product.

PTMG and TMP molecules were mixed in a 3:1 molar ratio. The fraction of AD
was controlled to generate samples of different molecular weights.
The acid numbers at the end of the reaction were near zero, - 
signifying complete conversion. 
The intrinsic viscosity of the molecules so generated
show a transition from the linear-like behavior at the low molecular weights
to randomly branched behavior at the high molecular weights. The crossover
between this region was found at $M_X = 66000$~g/mol. The high value of $M_X$
compared to mass of the oligomers (PTMG) 
indicates that a large fraction of
the TMP molecules has one of the OH groups unreacted.

In our modeling, we consider that all AD molecules present react completely
with some OH group. As in the synthesis, we consider molar ratios
$\molptmg : \moltmp = 3 : 1$. 
To keep the number of free parameters to a minimum, we 
assume the same rate constants ($k_1$) of esterification for the OH groups on
PTMG oligomers and that on the unreacted TMP molecules. Once one of the OH groups
on the TMP molecule has reacted, it can affect the rate constant for the second 
OH group because of the small separation of the OH groups on the TMP
molecules. Thus we assume
different rate constants $k_2$ ($k_3$) for rate constants of esterification
of OH group on TMP provided one (two) of the OH groups has already reacted.
In what follows, we denote unreacted TMP molecules as TMP$_0$ and TMP
molecules with $n$ of their OH groups reacted as TMP$_n$. The kinetic equations
considered are 
\begin{eqnarray}
&\concrate{OH}{PTMG}  =&  -k_1  \concohptmg \concacid \;\; , \nonumber \\
&\concrate{TMP}{0}  =&  - 3 k_1 \conctmp{0} \concacid \;\; , \nonumber \\
&\concrate{TMP}{1}  =&   3 k_1 \conctmp{0} \concacid - 2 k_2 \conctmp{1} \concacid \;\; , \nonumber \\
&\concrate{TMP}{2}  =&  2 k_2 \conctmp{1} \concacid -  k_3 \conctmp{2} \concacid \;\; , \nonumber \\
&\concrate{TMP}{3}  =&   k_3 \conctmp{2} \concacid \;\; , \nonumber \\
&\concrate{COOH}{}  =&  - k_1 \concacid \times \nonumber \\
& \{ \concohptmg +& 3 \conctmp{0} + \frac{2 k_2}{k_1} \conctmp{1} + 
  \frac{k_3}{k_1} \conctmp{2} \} \;\;.
\label{eqn:rates}
\end{eqnarray}
Here, $\concohptmg$ and $\concacid$ refer to the concentrations  of unreacted OH groups on the
PTMG molecules and number of unreacted COOH groups on the acid respectively. 
\conctmp{m} refers to the concentration of TMP$_m$ species. We fix the acid
concentration as $\molad = f_a \molptmg$, where $f_a$ determines the extent of
stoichiometric imbalance and hence the extent of the reaction. The rate equations
are solved numerically with the initial condition $t=0$, $\concacid = 2 \molad$, $\conctmp{0} = \moltmp$,
$\concohptmg = 2 \molptmg$ and $\conctmp{m} = 0$ for m being 1,2 or 3. 
At the end of the reaction $\concacid = 0$. Thus scaling the concentrations
by \concacid, the procedure involves numerical integration from $\concacid = 1$
to $\concacid = 0$.

\begin{figure}[htbp]
\centerline{ \includegraphics[width=8cm,clip=true]{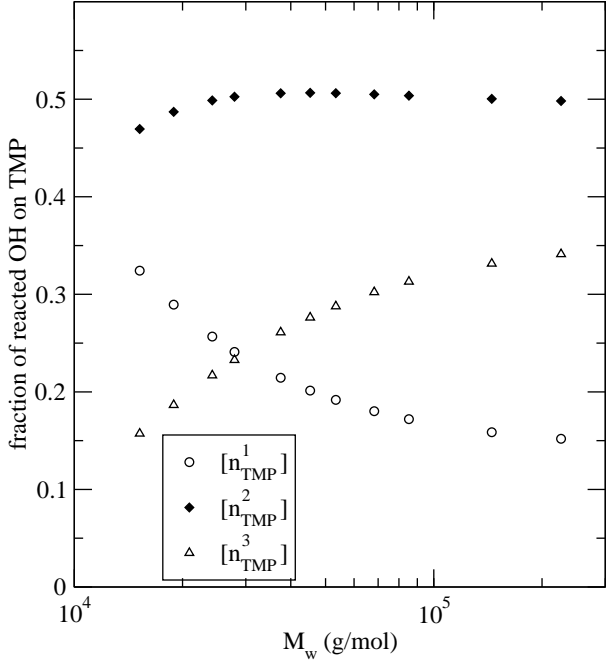}}
\caption{Fraction of TMP molecules having 1, 2 or 3 of the OH groups reacted 
as a function of average molar mass. $k_2 / k_1$ is assumed to be 0.8.}
\label{fig:TMPprob}
\end{figure}

To reduce the number of free parameters further, we assume that the presence of
one reacted OH group on a TMP molecule lowers the reaction rate by a certain 
fraction and the presence
of two reacted OH groups inhibit the reaction rate of the third
OH group independently ($k_3 = k_2^2$).
For a given acid concentration and value of $k_2/k_1$, the numerical
solution of the rate equations yields the probabilities of having different
reacted species in the final product (fig. \ref{fig:TMPprob}).
From these probabilities, we generate an ensemble of representative
molecules by first selecting
a species (PTMG, TMP$_n$ or AD)  with probabilities given by their respective weight 
fractions. Any unreacted acid group reacts with an OH group on either a
PTMG molecule or a TMP$_n$ molecule with
the probabilities from the solution of eqn.~\ref{eqn:rates}. For TMP$_n$
molecules, $n$ of the end groups are attached to AD molecules. For
PTMG molecules, ends are attached to an acid group to have the 
probability of reacted OH groups on PTMG the same as that given by
the solution of
eqn.~\ref{eqn:rates}. Since the initial species is selected on a
weight basis, in this way
we generate a weight-biased molecular distribution and 
the probability weight of each
individual molecule is simply the inverse of the total number of molecules
so generated.
For the rheological response, the molecules contribute with this probability
weight. We have here assumed that there are no ring molecules and that
the reactivities are independent of the size of the molecule - both
of which assumptions are expected to break down as the gel point is approached.
Also our analysis depends on the assumption of spatial homogeneity 
(continuously stirred reaction scheme).

\section{Static structure of branched PTMG polymers \label{sec:ptmg:static}}

\begin{figure}[htbp]
\centerline{ \includegraphics[width=8cm,clip=true]{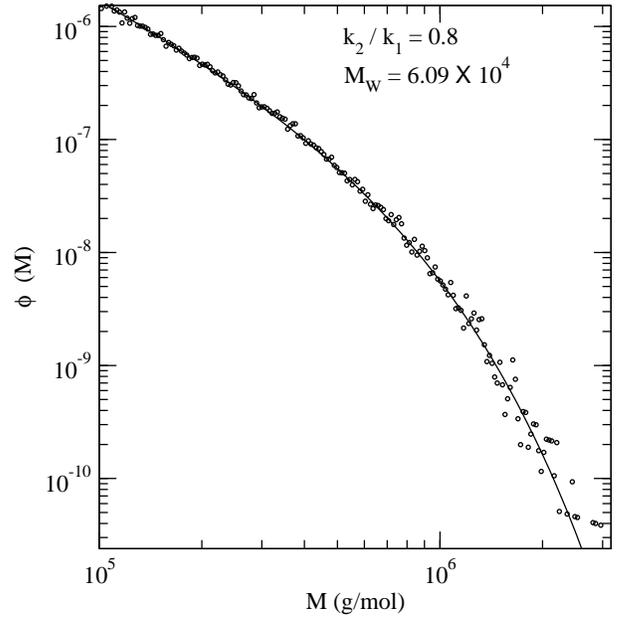}}
\caption{Determining $M_{char}$ : the line represents the best fit $\phi(M) \sim M^{-3/2} exp(-M/M_{char})$ 
indicating
the exponent $\tau = 5/2$ and $M_{char} = 2 \times 10^5$~g/mol. $k_2/k_1$ is fixed
to be $0.8$ in this plot.}
\label{fig:getmchar}
\end{figure}

For a given choice of $k_2/k_1$, the acid concentration $f_a$ is varied to
generate a series of different $M_W$ ensembles. For mean-field gelation
ensemble, the cut-off function in the molar mass distribution
$f(M/M_{char})$ in eq.~\ref{eq:massscaling} is explicitly known to be
\cite{rubinstein:colby}
\begin{equation}
f(M/M_{char}) = \exp\left[- \frac{M}{2 M_{char}} \right].
\end{equation}
Using this cut-off function and assuming that the exponent $\tau = 5/2$
in eqn.~\ref{eq:massscaling},
we use a two-parameter fit to determine $M_{char}$ from the tail region 
of the molar mass distribution (fig.~\ref{fig:getmchar}). Note that
because our ensemble is generated on weight basis, we fit a function
$\Phi(m) \sim M^{-3/2} \exp\left[- \frac{M}{2 M_{char}} \right]$
corresponding to  $\tau = 5/2$. The small-mass 
end of the distribution (not shown in the figure) does not conform 
to this form and shows
noisy features due to the finite size of the oligomers used
during synthesis.

\begin{figure}[htbp]
\centerline{ \includegraphics[width=8cm,clip=true]{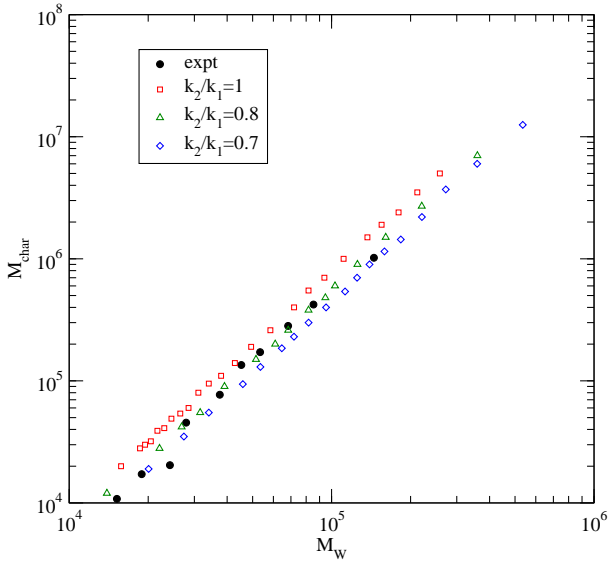}}
\caption{$M_{char}$ as a function of $M_W$ for several choices of $k_2/k_1$ ratios
superposed with the experimental data.}
\label{fig:mwmchar}
\end{figure}

In fig.~\ref{fig:mwmchar} we show $M_{char}$ as a function of $M_W$ for
three different choices of $k_2/k_1$. For $k_2 = k_1$, the simulation
values of $M_{char}$ are consistently higher than the experimental points, while
for $k_2 = 0.7 k_1$, the simulation values are consistently lower. For $k_2 = 0.8 k_1$,
the simulation results closely match the experimental values.

Ref.~\cite{lusignan:99} measured the intrinsic flow viscosity, $[\eta]$, and found a crossover
from linear-like behavior ($[\eta] \sim M^{0.8}$) at low molar mass to randomly
branched behavior  ($[\eta] \sim M^{0.45}$) at high molar mass. When 
all the samples of different $M_W$ were considered together, the crossover of
these two behaviors was found at $M_X = 6.6 \times 10^4$. Without 
knowledge of the detailed
interaction among the monomers, it is not possible to compute the intrinsic
viscosity. The intrinsic viscosity should depend linearly
with the radius of gyration in a good solvent - which again is beyond our 
ability to calculate. However, it is reasonable to assume that the radius of gyration
in a good solvent will be directly related to that in the $\Theta$ solution,
and in particular that the crossover molar mass between the linear-like
and the branched scaling will be the same. From the
numerical ensemble of the polymers, we used Kramers theorem\cite{rubinstein:colby}
 to calculate this
ideal radius of gyration.

\begin{figure}[htbp]
\centerline{ \includegraphics[width=8cm,clip=true]{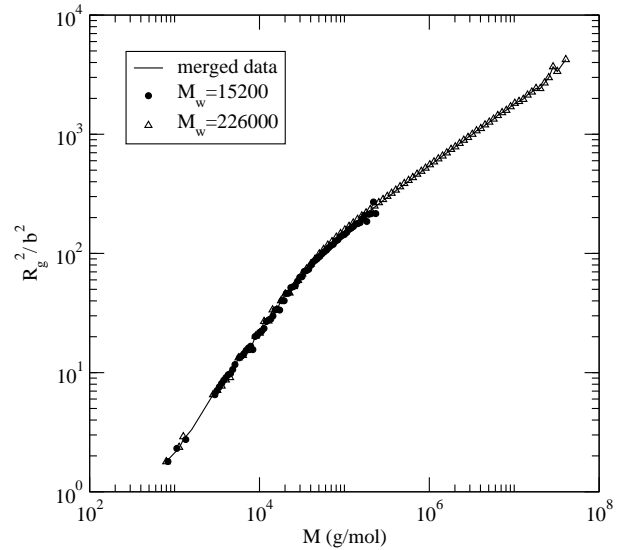}}
\caption{Ideal radius of gyration for $k_2 = 0.8 k_1$ and $M_W=15200$ (filled circles)
 and $M_W=226000$ g/mol (open triangles). The line  shows the combined behavior
of all the data sets from samples with different $M_W$ together.
$b$ is the Kuhn length and Kuhn mass is assumed to be $74$ g/mol.}
\label{fig:rgsplit}
\end{figure}

From the numerical ensemble of the molecules, we form histograms of molecules 
with respect to the mass of the molecules. For each bin in the histograms we 
calculate the average radius of gyration. For
these calculations, we assume that the Kuhn mass is $74$~g/mol 
\cite{lusignan:99} and the results are
in units of Kuhn length $b$. 
In fig.~\ref{fig:rgsplit} we plot the radius of gyration for two different molar
mass samples (symbols), both generated with $k_2/k_1 = 0.8$. Also shown is the
radius of gyration when all the different molar mass samples are considered together (line).
The individual $M_W$ ensembles roughly fall on this merged distribution line. 
A difference shows up only at the high molar mass limit - where the lower $M_W$ 
ensembles do not have any entries.

\begin{figure}[htbp]
\centerline{ \includegraphics[width=8cm,clip=true]{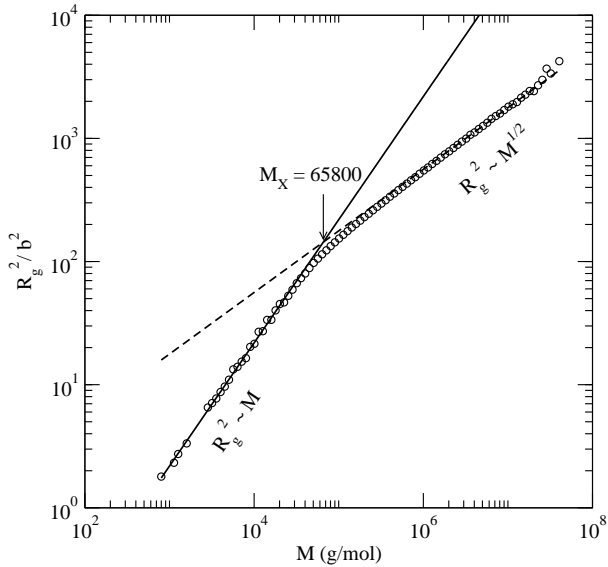}}
\caption{Crossover from linear to randomly branched behavior from radius 
of gyration. The symbols correspond to the radius of gyration determined by
making a histogram in mass from all the different $M_W$ samples considered
together. The solid line and the broken line corresponds to slopes
$1$ (linear) and $1/2$ (randomly branched) respectively. The arrow indicates
crossover at 65800 g/mol. $k_2/k_1$ is fixed at $0.8$.}
\label{fig:rgtot}
\end{figure}

In fig.~\ref{fig:rgtot} we plot the radius of gyration when all samples are considered
at $k_2 = 0.8 k_1$. At the low-molar-mass end, the data fits the form $R_g^2 \sim M$
corresponding to linear polymers. The high-molar-mass end fits the form
$R_g^2 \sim M^{1/2}$ corresponding to randomly branched polymers. The crossover
of these two behaviors is found by extrapolating the fits at $M_X = 65800$~g/mol.
Increasing (decreasing) the value of $k_2$ leads to   lowering (raising)
 $M_X$. For comparison, $k_2 = 1$ and $k_2 = 0.7$ respectively corresponds to
$M_X = 47500$~g/mol and $M_X = 89000$~g/mol. 
Since the same value of $k_2/k_1$ fits both this crossover behavior
and the variation of $M_{char}$ with $M_W$  with the experimental
results, only results with this value of $k_2/k_1$ are shown in the rest of
the paper.
The quantitative agreement with quite different experimental results using the
same parameter suggests that our simplistic assumptions about the reaction kinetics
are probably close to reality.

\begin{figure}[htbp]
\centerline{ \includegraphics[width=8cm,clip=true]{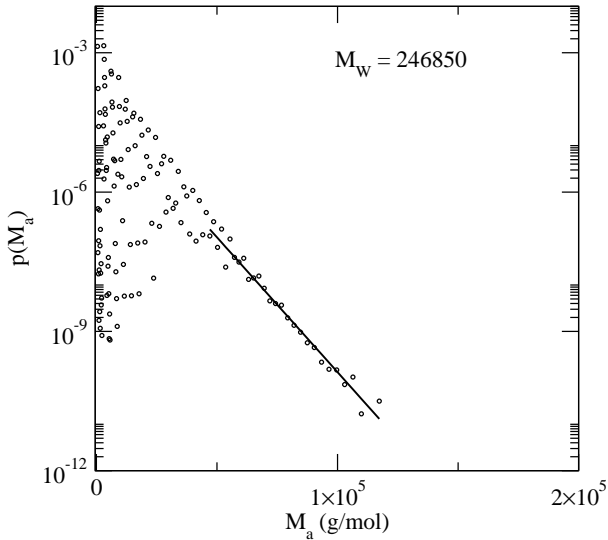}}
\caption{Probability distribution of linear segments at a fixed acid 
concentration. $p(M_{N,S})$ decays exponentially at large $M_{N,S}$.
The line denotes an exponential fit in the shown range.} 
\label{fig:floryfit}
\end{figure}

From the cross-over in radius of gyration, one might conclude that the typical
linear segment length is about 66000 g/mol (indeed, this is the conclusion
in \cite{lusignan:99} from the intrinsic viscosity data). 
With the detailed topological
connectivity at our disposal, we can probe the segment length between branch points 
in a different way. A linear segment can be made of PTMG oligomers connected by
acid groups alone - or with intervening TMP molecules with only two of the three OH
groups reacted. Such TMP connectors will have small side-arms (compared to entangled 
molecular mass) which will still behave as linear segments in rheological measurements.
We therefore  add the mass of such small side-arms to the backbone. This does not remove segments
which are too small to be rheologically important but still are connected at all ends
(for example, an H molecule with the cross bar formed by two TMP molecules will behave 
just like a four arm star). For this reason we take an alternative route than a
 simple average of the masses of the arms
(we believe that this alternative route is also rheologically most relevant).
Any random association to form linear segments attains a Flory distribution (most 
probable distribution) at mass scales much larger than the constituent elements.
The probability of having a segment of length $M_{a}$ is given by 
$p(M_{a}) = c \;\exp(-M/M_{N,S}) $,
with $c$ being an constant and, $M_{N,S}$ being the number-averaged molar mass
of the segments. From a histogram of linear segments, we determine $p(M_{a}$
and using an exponential fit at large $M_{a}$ determine the
number-averaged molar mass $M_{N,S}$ (fig.~\ref{fig:floryfit}). The 
weight-averaged molar mass for Flory distribution is twice $M_{N,S}$.

\begin{figure}[htbp]
\centerline{ \includegraphics[width=8cm,clip=true]{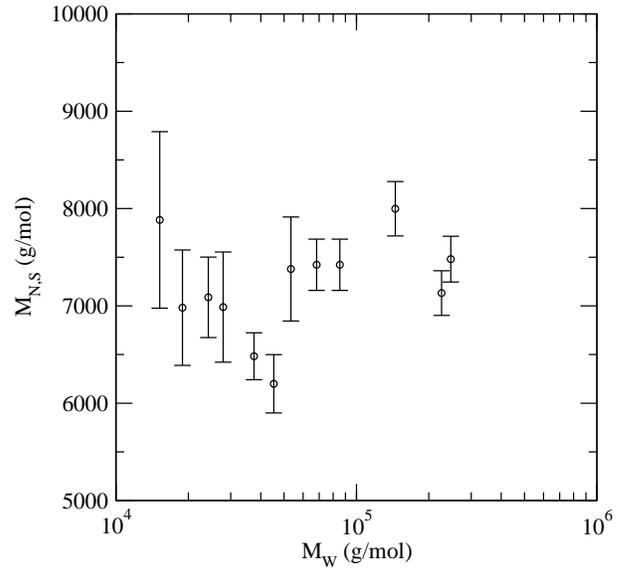}}
\caption{Estimate of number averaged molar mass of linear segments between branch points
as a function of $M_W$. The error bars are estimated from standard error in the exponential
fit of $p(M_a)$ (fig.~\ref{fig:floryfit}).}
\label{fig:mns}
\end{figure}

In fig~\ref{fig:mns} we plot $M_{N,S}$ as a function of $M_W$. The error bars correspond
to the error estimates in  exponential fit of $M(a)$ (fig.~\ref{fig:floryfit}).
determining the addition probability. 
At large $M_W$, the 
estimate of the linear segment length from this approach is $\sim 7200$~g/mol, 
which is almost an order smaller than $M_X$ determined from the crossover 
of radius of
gyration. This is due to the fact that lightly branched material like stars
or combs, which
dominate the mid-range in the mass distribution, have a radius of gyration which
is closer to that of linear polymers with the same molar mass 
than to that of randomly branched polymers.

\begin{figure}[htbp]
\centerline{ \includegraphics[width=8cm,clip=true]{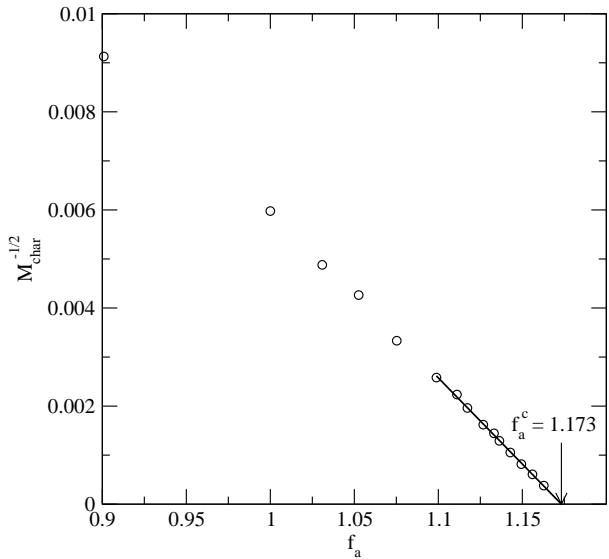}}
\caption{Dependence of characteristic mass $M_{char}$ on the acid fraction $f_a$. 
The line is a linear fit for high $M_W$ samples. The
intersection point with the zero x-axis gives the critical acid concentration $f_a^c$,
where the characteristic mass diverges.}
\label{fig:epsilon}
\end{figure}

As an estimate for closeness to the gelation transition, we define 
\begin{equation}
\epsilon \equiv \left(\frac{f_a^c - f_a}{f_a^c} \right),
\end{equation}
where, $f_a^c$ is the critical acid concentration where the
characteristic mass diverges.
Close to the gel point, the characteristic mass $M_{char}$ scales as
$M_{char} \sim \epsilon^{-2}$. Thus the plot of $1/\sqrt{M_{char}}$ 
versus $f_a$ shows a linear behavior for large $f_a$ (fig.~\ref{fig:epsilon}).
The point at which the line crosses the zero x-axis (infinite $M_{char}$)
determines $f_a^c = 1.173(1)$.
For $k_2/k_1 = 0.8$, the sample with average molar mass 220~Kg/mol corresponds to $f_a=1.156$
giving $\epsilon \sim 0.014$. 
The size of the largest branched molecule provides a characteristic length $\xi$ and
mean-field theory provides a self-consistency test by requiring that the molecules
of this characteristic size should overlap sufficiently. In 3 dimensions, this leads to
a critical value of the extent of the reaction $\epsilon_c \sim N^{-1/3}$ below which
the largest molecules no longer overlap significantly and the exponents change from
the mean-field results \cite{degennes:77,lusignan:99}. 
Taking the linear segment length 
$M_{N,S} = 7200$~g/mol
and Kuhn mass to be $74$~g/mol, there are on average approximately 100 Kuhn segments
between branch points. 
This estimate gives $\epsilon_c \simeq 100^{-1/3} \simeq 0.2$, so that the 
highest molar
mass samples are much closer to the gelation transition 
than the critical value and are therefore expected to show
non-mean-field behavior. We will meet the rheological consequences of
this critical behavior in the following.

\section{Rheological response of branched PTMG polymers \label{sec:ptmg:dyn}}
\subsection{Computational rheology \label{subsec:comprheo}}
To estimate the rheological behavior of the entangled branched PTMG polymers, we employ
a numerical approach \cite{larson:01,das:06}. For details, the reader is
referred to \cite{das:06}. We summarize the procedure qualitatively here
for completeness. The numerical approach is based on tube theory 
\cite{doi:book}, which
replaces the topological entanglements from neighboring chains by a hypothetical
tube surrounding a given chain. After a small strain, the stress is relaxed by 
the escape of the chains from the old tube constraints. This connects the stress
relaxation to the survival probability of the chains in their respective tubes. 
In a polymer melt, since all the polymers
are in motion, the tube constraint itself is not fixed over time. This constraint
release is handled by the dynamic dilation hypothesis 
\cite{marrucci:85,ball:89,colby:90}, which postulates a simple
relation between the tube diameter and the amount of unrelaxed material. 
A free end monomer relaxes part of its tube constraint at short times by 
constraint-release Rouse motion. 
At later times, the entropic potential,
which itself evolves due to constraint release, leads to a first-passage time approach \cite{milner:97}.
The contribution from a collapsed side arm is modeled by including increased friction
on the backbone, estimated from the time of collapse 
and the current tube diameter as a length-scale for diffusive hops from an Einstein relation.
For branch-on-branch architectures, the relaxation leads to a multi-dimensional 
Kramers' first-passage problem. We simplify this by recasting it to an effective
one-dimensional problem which has the required Rouse scaling, respects topological
connectivity and gives correct result at some special known 
limits \cite{das:06}. A linear or
effectively linear (branched material with collapsed side arms) chain can relax 
by reptation. When a large amount of material relaxes quickly, 
such that the dynamically dilated tube increases in diameter faster
than the rate permitted by Rouse relaxation,
the effective orientational constraint responds more slowly by constraint release
Rouse motion and the dynamic dilation is modified \cite{viovy:91,milner:98}. 
In addition, we include 
contributions from the Rouse motion inside the tube and fast forced redistribution 
of material at the early stages of the relaxation \cite{likhtman:02}. 

In computational rheology, starting from a numerical ensemble of
molecules, tube survival probability in discrete (logarithmic) time
is followed after an imaginary step strain 
\cite{larson:01,park:05a,park:05b,das:06}.
At each of these time steps, the amount
of unrelaxed material $\phi_t$ and the effective amount of tube constraint $\phi_{ST}$
is stored. Since the visco-elastic polymers have a very broad spectrum of relaxation,
we assume that the amount of material relaxed in each time
step contribute 
as independent modes in the stress relaxation modulus $G(t)$. Thus after all of the
molecules have relaxed completely, $G(t)$ is calculated as a sum over all these
independent modes. The complex modulus $G^* (\omega)$ 
at frequency $\omega$ is defined by
\begin{equation}
G^* (\omega) = i \omega \int_0^{\infty} G(t) exp(-i \omega t) dt.
\end{equation}
The real and imaginary parts of $G^* (\omega)$, storage modulus $G^{'}(\omega)$ and
the dissipative modulus $G^{''}(\omega)$ respectively, are of particular interest since
they are measured in oscillatory shear experiments. The zero-shear viscosity $\eta$
is calculated from
\begin{equation}
\eta_0 = \lim_{\omega \rightarrow 0}
\frac{G^{''}(\omega)}{\omega},
\end{equation}
and the steady-state compliance $J_e^0$ is calculated from
\begin{equation}
J_e^0 =   \lim_{\omega \rightarrow 0} \frac{G^{'}}{(G^{''})^2}.
\end{equation}
All the integrations are replaced by sums over discrete timesteps of the relaxation.

The calculations have a few free parameters. The material-dependent parameter of entanglement
molar mass $M_e$ is related to the plateau modulus $G_0$ by \cite{larson:03}
\begin{equation}
G_0 \equiv \frac{4}{5} \frac{\rho R T}{M_e},
\end{equation}
where, $\rho$ is the polymer density and $T$ is the temperature. The timescale is
set by the entanglement time $\tau_e$ which is the Rouse time of the
chain segment between entanglements. When the molar mass of the segments is scaled by $M_e$
and the time is scaled by $\tau_e$, in the approximation of tube theory,
polymers of the same topology but of different chemical composition relax the same
way. We assume that, for a side-arm relaxing completely at certain time $t_a$, at
times much larger than $t_a$, the motion of the associated branch-point can be 
modeled as a simple diffusion process with hop size $p a$ at the timescale of
$t_a$. Here $a$ is the tube diameter and $p$ is a numerical factor. We use 
$p^2 = 1/40$ as used in \cite{das:06} to fit a wide range of different 
experimental data. The dynamic dilation hypothesis assumes that the 
effective tube diameter depends on the amount of unrelaxed material and
the effective number of entanglements associated with a segment of length Z
scales as 
$Z \rightarrow \phi_t^{-\alpha}$. We choose the dynamic dilation exponent $\alpha= 1$ 
in our calculations.

For the class of polymers 
considered in this study, the number of branches on a given molecule can be 
quite large. Also a large number of molecules need to be considered to ensure
that a single massive molecule does not affect the results disproportionately.
In our approximations, the relaxation of the different molecules are coupled
only via the amount of unrelaxed material $\phi_t$. This enables us to
divide the ensemble of molecules in several subsystems and follow the 
relaxation process independently at each step, communicating the local $\phi_t$
to other processors at the end of each time step. The minimal communication
needed makes the  parallel-code scale almost perfectly with the number of
processors and most importantly allows us to probe closer to the
gelation transition, where the memory requirement becomes larger than available
on single processors. The source code, precompiled executables and documentation of 
the program are available from \verb+http://sourceforge.net/projects/bob-rheology+.

\subsection{Dynamic exponents for branched PTMG \label{subsec:ptmg:expon}}
To calculate the dynamic properties  of the branched PTMG molecules, 
we generated ensembles of $5 \times 10^5$ molecules at each
$M_W$ considered and followed the relaxation after a small step-strain.
In the absence of high-frequency measurements for this material,
we take $M_e$ and $\tau_e$ as free parameters - fitted to describe
the dynamic properties. Without the complications of occasional 
ester groups from the esterification and the butyl side
groups from the TMP molecules, the present polymers resemble 
polytetrahydrofuran (PTHF). For PTHF, treating it as
an alternating copolymer of ethylene and ethylene oxide,
the estimate of the entangled 
molecular weight is $M_e \simeq 1420$~g/mol \cite{fetters:pv}.
We use this value as our rough first guess for $M_e$ and fix $\tau_e$
by matching the zero-shear viscosity with experimental results at 
the intermediate molar mass range of the experimental data.

\begin{figure}[htbp]
\centerline{ \includegraphics[width=8cm,clip=true]{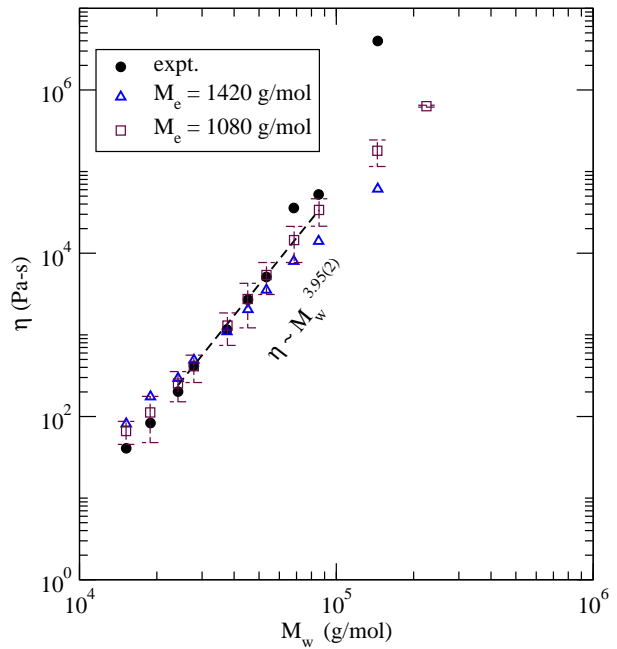}}
\caption{Zero-shear viscosity $\eta$ as a function of weight-average
molar mass. Filled circles are the experimental data from \cite{lusignan:99}.
The triangles and the squares are the results from our calculations
using $M_e = 1420$ and $1080$~g/mol respectively. The error bars for
the $1080$~g/mol calculations represent the effect of uncertainty
in $\eta$ corresponding to the experimental uncertainty of $10\%$
in determining $M_W$. The line is a power-law fit to the
$M_e = 1080$~g/mol results in the 
intermediate mass region with the exponent 3.95.}
\label{fig:visc}
\end{figure}

In fig.~\ref{fig:visc} we plot the zero shear viscosity $\eta$ for
different values of $M_W$. The filled circles are the experimental
results from \cite{lusignan:99}. The triangles are results
from our calculations with $M_e = 1420$~g/mol and 
$\tau_e = 1.8e-7$~s. For this choice of $M_e$, the viscosity
increases slowly with $M_W$ compared to the experimental data.
The squares are results with $M_e = 1080$~g/mol and
$\tau_e = 3.45e-8$~s. This choice of $M_e$ is able to reproduce
the experimental viscosity data over half a decade in $M_W$.
A power-law fit in the intermediate range (shown as a dashed 
line in the figure) gives the viscosity exponent $s = 3.95(2)$.
In the experiments, there is an uncertainty of $10\%$ in determining
$M_W$. The error bars in the $M_e = 1080$~g/mol data show the
associated uncertainty in viscosity.
At the largest $M_W$, our calculations and the experimental
data show opposite trends. The viscosity from our calculations
shows a trend of lowering of the exponent at the largest $M_W$,
while the experimental data shows a sharp increase. For
rest of the results in this section, we use $M_e = 1080$~g/mol.

\begin{figure}[htbp]
\centerline{ \includegraphics[width=8cm,clip=true]{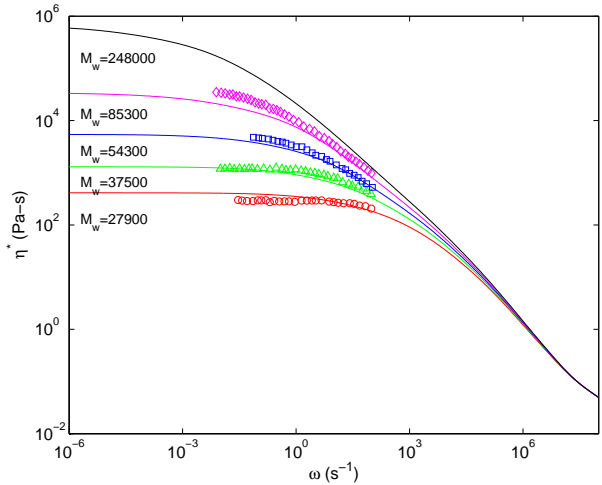}}
\caption{Complex viscosity $\eta^* (\omega)$ for selected $M_W$.
Lines are from calculations in this study, symbols are experimental
data from \cite{lusignan:99}.}
\label{fig:comvis}
\end{figure}

\begin{figure}[htbp]
\centerline{ \includegraphics[width=8cm,clip=true]{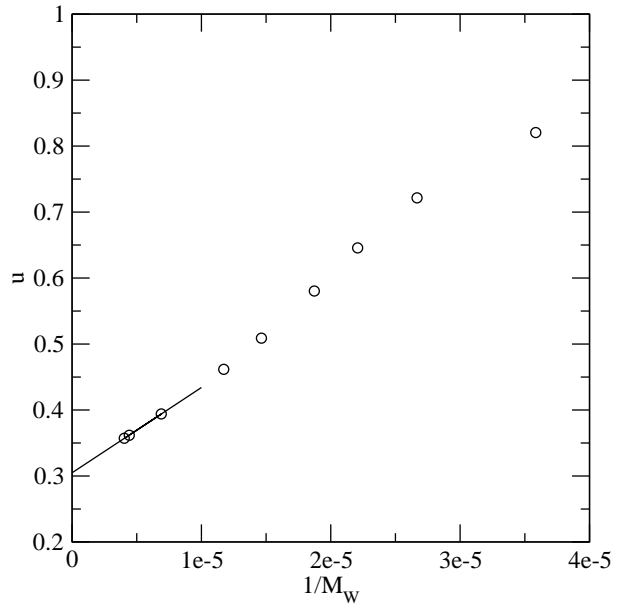}}
\caption{Variation of the relaxation exponent $u$ with $M_W$. 
The line shows the extrapolation used for large $M_W$ to find
the value of $u$ at the gel transition.}
\label{fig:ufit}
\end{figure}

Fig.~\ref{fig:comvis} shows the frequency dependence of the
complex viscosity $\eta^*(\omega)$ for several different values of $M_W$. Symbols
represent experimental data from \cite{lusignan:99} in
the intermediate mass range, where zero-shear viscosity 
from our calculation matches with the experimental values. 
$\eta^*(\omega)$ shows an approximate power-law behavior 
with exponent $1-u$. Away from the gelation transition,
this power-law window is limited. We fitted power laws 
in the frequency range 10-100~s$^{-1}$ to estimate
the exponent. Since $1/M_W \sim \epsilon$, we plotted
$u(M_W)$ from such fitting as a function of $1/M_W$
(fig.~\ref{fig:ufit}). Linear extrapolation to
$1/M_W \rightarrow 0$ gives the limiting value of 
$u = 0.305(1)$, which corresponds closely to the
experimental value of $u=0.31(2)$.

\section{Mean-field gelation ensemble \label{sec:gelation}}

The segment length for branched PTMG considered in the earlier part of the 
paper is largely determined by the size of the oligomer used to synthesize
the polymers. In this section we turn to a hypothetical series of polymers
which fall in the category of mean-field gelation class. We consider 
linear molecules of type $A$ which are Flory distributed and tri-functional
groups $B$ which have zero mass. The molecules are formed by the rule that
$A$ reacts with $B$ and neither  $A$-$A$ nor $B$-$B$ reactions are allowed.
Furthermore, we assume that, at the end of the reaction, all $B$ bonds are
attached to some $A$ molecules. Thus, as in the case of branched PTMG polymers,
extent of the reaction is determined by stoichiometric mismatch. As before,
we assume that there are no closed loops. The final distribution of the 
molecules are described by only two parameters: $M_{N,S}$, number-averaged 
molar mass of the linear segments and $p_b$, the 
branching probability. The molecules are generated by selecting
the first strand with a Flory distribution of length $M_{N,S}$ and
adding Flory distributed branches recursively on both ends with probability $p_b$. 

The static properties of these molecules can be solved analytically. 
The characteristic molar mass diverges when $p_b$ is 0.5 
($\epsilon \equiv (0.5 - p_b)/0.5$). Using
seniority variables to approximately describe the
hierarchical relaxation,
ref.~\cite{read:01} found that the entangled contribution
to the terminal relaxation time of this class of polymers does not diverge 
at the percolation threshold. Their calculation did not include the 
constraint release Rouse
modes, contributions from which will still be divergent in the absence of
a diverging entangled contribution. For the calculations presented in this
section, we assume $M_e = 1120$~g/mol and $\tau_e = 1.05 \times 10^{-8}$~s,
corresponding to high-density polyethylene at 150$^o$~C \cite{das:06}.
For each value of $p_b$ and $M_{N,S}$ considered, we generate an ensemble of
$2 \times 10^5$ molecules and follow the relaxation after a step strain.
To estimate statistical errors involved in our calculations, for each
case we repeat the calculation 3 times with different sets of molecules 
(generated by different random seeds).

\begin{figure}[htbp]
\centerline{ \includegraphics[width=8cm,clip=true]{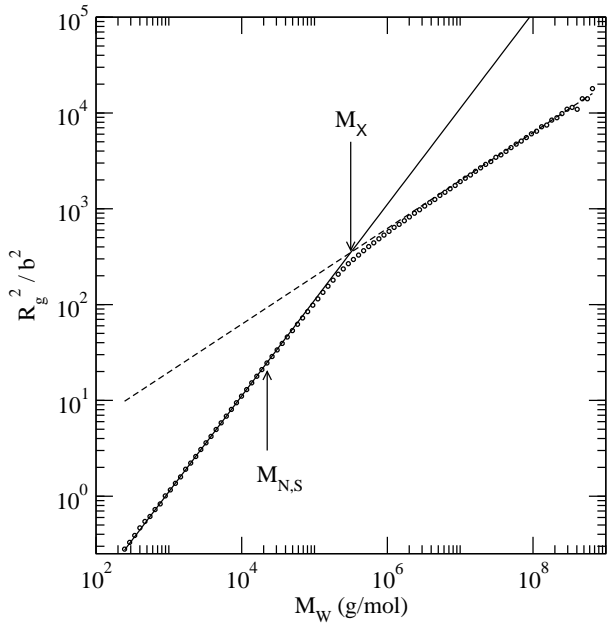}}
\caption{Radius of gyration for gelation ensemble with $M_{N,S} = 22400$~g/mol
with $p_b = 0.49$.
The cross-over mass $M_X = 314400$ is roughly 14 times larger than 
$M_{N,S}$.}
\label{fig:gelrg}
\end{figure}

In fig.~\ref{fig:gelrg} we plot the mass dependence of the radius of gyration
for $p_b = 0.490$ ($\epsilon = 0.02$) and segment length $M_{N,S} = 22400$~g/mol
(number of entanglements between branch points $N/N_e = 20$). 
As in the case of branched PTMG, the extrapolated cross-over 
mass $M_X$ in radius of gyration  from linear to the randomly branched behavior
is much larger than $M_{N,S}$. Because the segments are Flory distributed
with out a lower cutoff, the difference is even larger in this case 
($M_X / M_{N,S} \sim 14$).

\begin{figure}[htbp]
\centerline{ \includegraphics[width=8cm,clip=true]{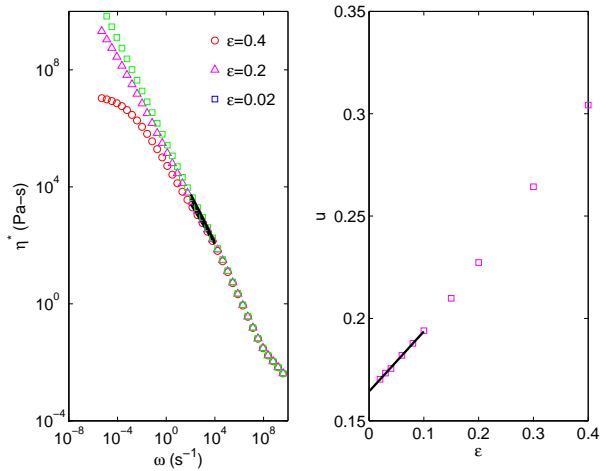}}
\caption{Estimating the exponent u : the left panel shows $\eta^* (\omega)$
for $N/N_e = 20$ and $\epsilon=0.4$, $0.2$, and $0.02$. Data for
different $\epsilon$ are fitted separately to a power law with exponent 
$1-u$ in the frequency range $10^{2} - 10^4$~s$^{-1}$. 
The exponent
$u$ so obtained, as a function of $\epsilon$,  is plotted in the right 
panel. A linear fit for $\epsilon \le 0.1$ was used to find the
limiting value of $u$ at this $N/N_e$. }
\label{fig:gelfitu}
\end{figure}

Fig.~\ref{fig:gelfitu} illustrates the procedure followed for estimating
the apparent relaxation exponent $u$
(when the relaxation dynamics are entangled there is no reason to
expect true power-law behavior, but an apparent power law can hold
as a good approximation for a sizeable range of relaxation timescales
\cite{rubinstein:90}). The left subpanel shows the variation of
the complex viscosity $\eta^*$ with frequency for three different $\epsilon$
for $N/N_e = 20$. At the lowest frequencies, for $\epsilon = 0.4$, the terminal
relaxation leads to significant deviation from the power-law behavior. For
smaller values of $\epsilon$, this deviation shifts to smaller frequencies.
For different $\epsilon$, we fit a power law with exponent $1-u$ in the
frequency range $10^{2} - 10^4$~s$^{-1}$. In the right subpanel of
fig.~\ref{fig:gelfitu} we plot the dependence of such apparent $u$ with
$\epsilon$. For $\epsilon \le 0.1$, the values of $u$ shows a linear 
dependence on $\epsilon$. A linear fit was used to estimate the
extrapolated value of $u$ at $\epsilon = 0$.

\begin{figure}[htbp]
\centerline{ \includegraphics[width=8cm,clip=true]{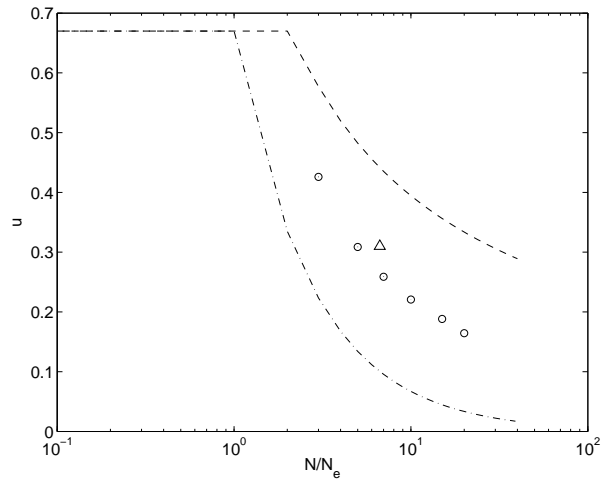}}
\caption{Variation of (apparent) relaxation exponent $u$ with $N/N_e$. 
The circles
are the data obtained from our calculations. The dotted line is the 
prediction of hierarchical relaxation model of \cite{rubinstein:90}
and the dashed line is the phenomenological form used in \cite{lusignan:99}.
The triangle corresponds to the exponent $u$ for the branched PTMG polymers,
when the segment length is estimated by fitting a Flory distribution to the
probability distribution of segment lengths. }
\label{fig:gelexpu}
\end{figure}

Fig.~\ref{fig:gelexpu} shows the extrapolated values of $u$ (circles) at $\epsilon = 0$ from
the procedure outlined in fig.~\ref{fig:gelfitu} as a function of number
of entanglements between branch points $N/N_e$. The error estimates for $u$
are smaller than the size of the symbols. An approximate calculation
of hierarchical relaxation in entangled mean-field gelation tube model at
the gel point \cite{rubinstein:90} predicted a form of $u$ as
\begin{equation}
u = \psi \frac{N_e}{N} \;\;\; {\text{for}} \;\; N \ge N_e,
\label{eq:u:rubenstein}
\end{equation}
with $\psi$ being a constant. Both theory and experiments for
$N < N_e$, where unentangled Rouse dynamics dominate, suggest 
$\psi \approx 0.67$.  The dotted line shows the prediction
from eqn.~\ref{eq:u:rubenstein}. 
Ref.~\cite{lusignan:99} uses an empirical function,
\begin{equation}
u = \left\{ \begin{aligned} &0.67  & N < 2 N_e \\
                    &\frac{3}{3 + 2 \ln(N/N_e)} \;\;& N > 2 N_e, 
       \end{aligned}
      \right.
\label{eq:u:lusignan}
\end{equation}
to describe the dependence of $u$ on $N/N_e$ from experimental data.
The dashed line in fig.~\ref{fig:gelexpu} shows this phenomenological
function. Results from our calculations fall roughly midway
between the prediction of the approximate model \cite{rubinstein:90} and the
phenomenological fit to data in \cite{lusignan:99}.  The significant deviation from
eq.~\ref{eq:u:lusignan} is mostly because \cite{lusignan:99}
uses $M_X$ as an indicator for $M_{N,S}$ (so overestimating it)
and to some extent because
$u$ changes appreciably as the limit $\epsilon \rightarrow 0$ is
considered (fig.~\ref{fig:gelfitu}). When plotted against our
estimate of linear segment length (shown as triangle in fig.~\ref{fig:gelfitu}),
the exponent $u$ corresponding to the experiments reported in
ref.~\cite{lusignan:99} matches closely with our calculations on
gelation ensemble.

\begin{figure}[htbp]
\centerline{ \includegraphics[width=8cm,clip=true]{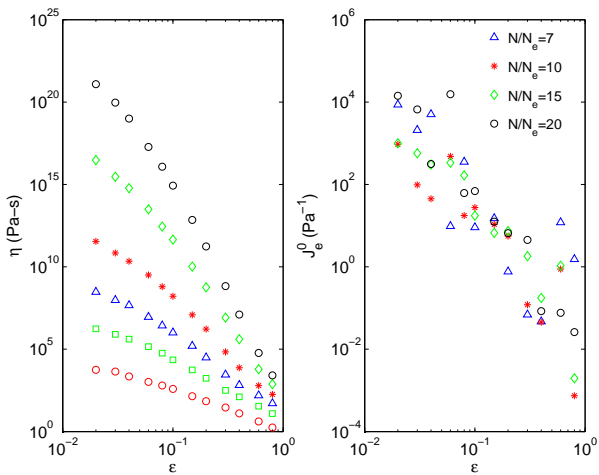}}
\caption{Zero-shear viscosity $\eta$ and recoverable compliance $J_e^0$ as
a function of $\epsilon$ for different $N/N_e$. The left subplot shows
$\eta$ for $N/N_e = 3$, $5$, $7$, $10$,  and $20$. The right subplot
shows $J_e^0$ for only $N/N_e \ge 7$.}
\label{fig:gelvisje0}
\end{figure}

The left subpanel of fig.~\ref{fig:gelvisje0} shows the zero-shear viscosity $\eta$
as a function of $\epsilon$ for different value of $N/N_e$ (larger $N/N_e$ data have
higher viscosity at the same $\epsilon$). The right subpanel shows the
recoverable compliance $J_e^0$ for $N/N_e = 7$, $10$, $15$, and $20$.
The data shows a large amount of scatter.  $J_e^0$ can be expressed as the 
first moment of $G(t)$ : $J_e^0 = \left(\int_0^{\infty} t G(t) dt \right) / \eta^2$.
Thus, $J_e^0$ is particularly susceptible to the long-time decay
of $G(t)$. The longest relaxation time is dominated by just a few
of the high molar mass molecules in our ensemble. Thus the variation of
$J_e^0$ with the particular ensemble considered is large.
For small $N/N_e$, the relative contribution from this tail region of
molar mass distribution is even higher. For $N/N_e < 7$, the scatter
becomes larger than the value of $J_e^0$ and they are neither shown
in the figure nor considered for further analysis.

\begin{figure}[htbp]
\centerline{ \includegraphics[width=8cm,clip=true]{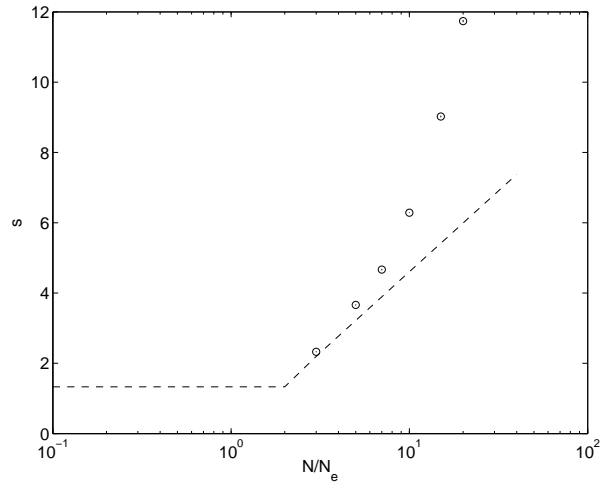}}
\caption{Viscosity exponent $s$ as a function of $N/N_e$. The circles are
results from our calculations and the dashed line is the phenomenological form of
Lusignan {\em et~al.\ }\cite{lusignan:99}.}
\label{fig:gelexps}
\end{figure}

At the smallest values of $\epsilon$ plotted in the log-log plot in fig~\ref{fig:gelvisje0}, 
in double log plot,
the slope of $\eta$ with $\epsilon$ starts to decrease. To probe at 
even smaller $\epsilon$ would require much larger computations
than used in this study. Instead we focus our attention in the
range of $\epsilon$ between $0.006$ and $0.2$, where the viscosity
for all values of $N/N_e$ shows approximate power-law dependence on
$\epsilon$. Fig.~\ref{fig:gelexps} shows the viscosity exponent
$s$ as a function of $N/N_e$. Also shown is the phenomenological
form of \cite{lusignan:99} as dashed line
\begin{equation}
s = \left\{ \begin{aligned} & 1.33  & N < 2 N_e \\
                    & 2 \ln(N/N_e) \;\;& N > 2 N_e. 
       \end{aligned}
      \right.
\label{eq:s:lusignan}
\end{equation}
Results from our calculations show a much sharper increase of $s$ with
$N/N_e$ than predicted by this functional form.

\begin{figure}[htbp]
\centerline{ \includegraphics[width=8cm,clip=true]{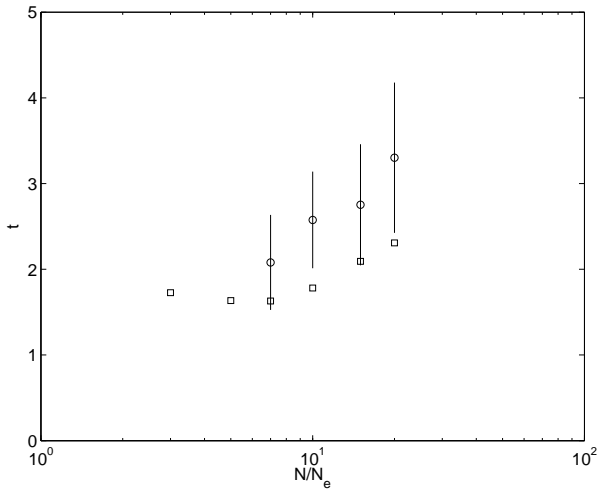}}
\caption{Recoverable compliance exponent t as a function of $N/N_e$. The circles
represent results from a direct power-law fit of the form $J_e^0 \sim \epsilon^{-t}$.
The error bars are estimated from the variance in $J_e^0$ from three independent
sets of molecular ensemble. The squares are calculated by using the hyperscaling 
relationship (eq.~\ref{eq:hyperscaling}). For $N/N_e = 3$ and $5$, the scatter
in $J_e^0$ is too large for a direct estimate of $t$.}
\label{fig:gelexpt}
\end{figure}

In fig.~\ref{fig:gelexpt}, we plot the recoverable compliance exponent $t$
(circles) as a function of $N/N_e$. Because of large scatter in $J_e^0$,
the error estimates in this case are large (error bars are estimates of
error from the variance obtained from three independent sets of calculations).
Since both $\eta$ and $J_e^0$ can be expressed as integrals over $G(t)$, the exponents
$u$, $s$ and $t$ are not independent. If $G(t)$ behaves like $t^{-u}$ till
the longest relaxation time, one gets a dynamic hyperscaling relationship
among the exponents
\begin{equation}
u = \frac{t}{s + t}.
\label{eq:hyperscaling}
\end{equation}
The estimates of $t$ from estimates of $u$ and $s$ using this hyperscaling relation 
is shown as the squares in fig.~\ref{fig:gelexpt}. In the range of $N/N_e$,
where we have direct estimates of $t$, estimates from the hyperscaling
relationship falls below the direct estimate.

\begin{figure}[htbp]
\centerline{ \includegraphics[width=8cm,clip=true]{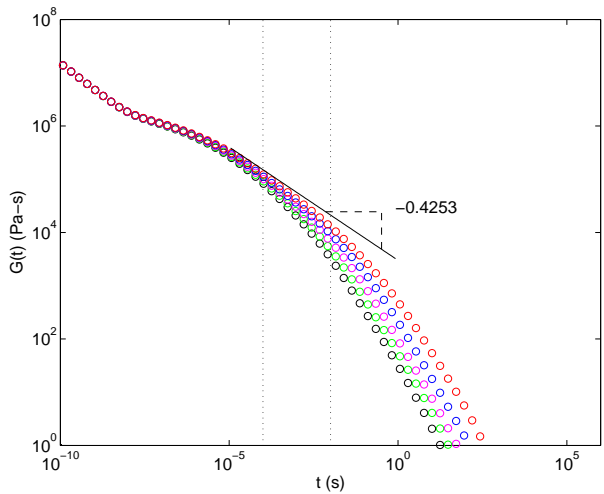}}
\caption{G(t) (symbols) for $N/N_e = 3$ for $\epsilon = 0.1$, $0.08$, $0.06$, $0.04$ and
$0.02$. As $\epsilon$ is made smaller, $G(t)$ decays slowly. The vertical dotted
lines show the range in which the complex viscosity is fitted (in frequency) to find
the value of the exponent $u$. The solid line represents this limiting power 
law decay $G(t) \sim t^{-u}$ with the indicated slope $u=0.4253$.}
\label{fig:gelgt}
\end{figure}

To explore why this is so,
in fig.~\ref{fig:gelgt} we plot the decay of $G(t)$ for different values of $\epsilon$
and $N/N_e = 3$.
Also shown is the limiting power law decay suggested from fitting the complex viscosity
data. For even the lowest $\epsilon$ studied here, the power-law behavior holds in only
a small window of the relaxation time and the contribution in $\eta$ or $J_e^0$ from
decay at times larger than this power-law window is not negligible. Thus, the dynamic
hyperscaling relationship holds only approximately (fig.~\ref{fig:gelexpt}).

\section{Conclusions \label{sec:concl}}
We have presented a simple kinetic modeling scheme for the gelation ensemble
polymer synthesis in \cite{lusignan:99}. With just one global
fitting parameter describing the branching chemistry, 
we are able quantitatively to reproduce the variation of
characteristic molar mass $M_{char}$ as a function of $M_W$ and
the behavior of intrinsic viscosity as a function of molar mass.
With the detailed knowledge of the molecular topology in
our calculations, our estimate of the average segment length 
between branch-points is much lower than estimated in \cite{lusignan:99}. 
We have used a numerical technique based on the tube theory
of polymer melts to calculate the dynamic response of the polymers
in the linear response regime. For intermediate ranges of $M_W$,
both the complex viscosity $\eta^* (\omega)$ and the zero-shear
viscosity $\eta$ matches with the experimental findings. For
the largest $M_W$ considered, $\eta$ in our calculations is
significantly lower than the experimental data. At those $M_W$,
our estimate of closeness to the gelation transition $\epsilon$
is well below the Ginzburg-de~Gennes criterion \cite{degennes:77} which is
a feature of the highest $M_W$ polymers that distinguishes them
from the others in the set. Hence the 
difference is likely to be due to non-mean-field behavior of these
samples.  Also, the four highest molar mass samples were 
prepared under  slightly different conditions - where a partial
reaction was carried out with stirring and, for the later
part, the samples were reacted without stirring at a slightly
elevated temperature \cite{lusignan:99}. Thus, our assumption
of continuous stirred reaction may not be completely true for
these samples. 
The dynamic exponents calculated from our calculations match
with the experimental findings in the relevant $M_W$ range.

To investigate the behavior of the dynamic exponents with
average segmental lengths between branch points, we calculated
the relaxation properties of a series of molecules in the
ideal mean-field gelation ensemble. The dependence of the
relaxation exponent $u$ on $N/N_e$ falls about midway between the
prediction of \cite{rubinstein:90} and the phenomenological 
form of \cite{lusignan:99}.
We find that the viscosity exponent becomes smaller as 
$\epsilon$ is lowered. This is due to the dominance of
supertube Rouse relaxation at long time scales for this
class of polymers and, for small enough $\epsilon$, the
viscosity exponent for any $N/N_e$ approaches the Rouse value 
applicable to the unentangled polymers. 

The recoverable compliance exponent $t$ in our calculations
have values similar to those found in experiments. It is worth noting,
however, that the
magnitude of $J_e^0$, when calculated from our algorithm,
is found to be much larger than
experimental values on similarly branched systems. Being
the first moment of the relaxation modulus $G(t)$, the
dominant contribution to $J_e^0$ comes from the long time
behavior of $G(t)$, so is very sensitive to the assumptions
on which the relaxation dynamics of the very largest clusters
in the ensemble is based. The computational scheme we used to follow the
relaxation in the melt
extrapolates ideas of dynamic dilation and supertube relaxation
which originally were formulated for linear or lightly branched
systems to a highly branched system. In particular it assumes that
the final supertube relaxation follows a Rouse scaling
corresponding to a linear object. The final relaxation,
provided that the tail of the distribution is long enough, of
largely unentangled high molar mass molecules may find a
faster route by showing a Zimm like relaxation, by which the
largest clusters relax hydrodynamically in an effective
solvent provided by the smaller clusters. In linear systems
the transition molecular weight for this is the same as
that for incomplete static screening of the larger molecules'
self-interactions.
Experimental results on model systems with high seniority and well characterized
branching and molar mass are needed to quantitatively
test the validity of the theory for accounting the
long-time decay of stress in such highly branched systems.

In summary, a numerical calculation of the entangled
rheology of a series of mean-field gelation ensemble
polymers provide a remarkable support of the
accuracy of the hierarchical relaxation process
suggested by the tube model.

\section*{acknowledgments}
The authors gratefully acknowledge communications with R.~Colby and
C.~P.~Lusignan. We thank L.~J.~Fetters for providing the value of
$M_e$ for PTHF. Funding for this work was provided by EPSRC.


\begin{thebibliography}{99}
\bibitem{lusignan:99} C.~P.\ Lusignan, T.~H.\ Mourey, J.~C.\ Wilson, and R.~H.\ Colby,
Phys. Rev. E {\bf 60}, 5657 (1999).
\bibitem{das:06} C.\ Das, N.~J.\ Inkson, D.~J.\ Read, M.~A.\ Kelmanson, and 
T.~C.~B.\ McLeish, J. Rheol., {\bf 50}, 207 (2006).

\bibitem{stauffer:92} D.\ Stauffer and A.\ Aharony, {\em Introduction
to Percolation Theroy}, 2nd ed. (Taylor and Francis, London, 1992).
\bibitem{degennes:79} P.~G.\ de~Gennes, {\em Scaling Concepts in Polymer
Physics} (Cornell University Press, Ithaca, 1979).
\bibitem{rubinstein:colby} M.\ Rubinstein and R.~H.\ Colby,
{\em Polymer Physics}, (Oxford University Press, Oxford, 2003).

\bibitem{stockmayer:43} W.~H.\ Stockmayer, J.\ Chem.\ Phys. {\bf 11}, 45 (1943).
\bibitem{flory:53} P.~J.\ Flory, {\em Principles of Polymer Chemistry}
(Cornell University Press, Ithaca, 1953).
\bibitem{alder:90} J.\ Alder, Y.\ Meir, A.\ Aharony, and A.~B.\ Harris,
Phys.\ Rev.\ E {\bf 41}, 9183 (1990).
\bibitem{valles:79} E.~M.\ Valles and C.~W.\ Macosko, 
Macromolecules {\bf 12}, 521 (1979).
\bibitem{stauffer:83} D.\ Stauffer, A.\ Coniglio, and M.\ Adam,
Adv.\ Polym.\ Sci.\ {\bf 44}, 103 (1983).
\bibitem{durand:87}
D.\ Durand, M.\ Delsanti, M.\ Adam, and J.~M.\ Luck, Europhys.\ Lett.\
{\bf 3}, 297 (1987).
\bibitem{winter:87}
H.~H.\ Winter, Prog.\ Colloid Polym.\ Sci.\ {\bf 75}, 104 (1987).
\bibitem{martin:89}
J.~E.\ Martin, D.~A.\ Adolf, and J.~P.\ Wilcoxon, Phys.\ Rev.\ A
{\bf 39}, 1325 (1989).
\bibitem{nicol:01}
E.\ Nicol, T.\ Nicolai, and D. Durand, Macromolecules, {\bf 34},
5205 (2001).
\bibitem{gasilova:02}
E.\ Gasilova, L.\ Benyahia, D.\  Durand, and T.\ Nicolai,  
Macromolecules, {\bf 35}, 141 (2002).
\bibitem{cates:85} M.~E.\ Cates, J.\ Phys.\ France {\bf 46}, 1059 (1985).

\bibitem{rubinstein:90} M.\ Rubinstein, S.\ Zurek, T.~C.~B.\ McLeish, and
R.~C.\ Ball, J.\ Phys.\ France {\bf 51}, 757 (1990).

\bibitem{doi:book} M.\ Doi, and S.~F.\ Edwards,  {\em The Theory of Polymer 
Dynamics} (Claredon Press, Oxford, U.K., 1986).
\bibitem{marrucci:85} G.\ Marrucci, 
J.\ Polym.\ Sci., Polym.\ Phys.\ Ed.\ {\bf 23}, 159 (1985).
\bibitem{ball:89}
R.~C.\ Ball, and T.~C.~B.\ McLeish, Macromolecules {\bf 22},  1911 (1989).
\bibitem{colby:90}  R.~H.\ Colby, and M.\ Rubinstein,
Macromolecules {\bf 23},  2753 (1990).
\bibitem{viovy:91} J.~L.\ Viovy, M.\ Rubinstein and  R.~H.\ Colby,
Macromolecules {\bf 24}, 3587 (1991).
\bibitem{milner:98} S.~T.\ Milner, T.~C.~B.\ McLeish, R.~N.\ Young, 
 A.\ Hakiki, and J.~M.\ Johnson, Macromolecules, {\bf 31}, 9345 (1998).

\bibitem{degennes:77} P.~G.\ de~Gennes, J.\ Phys.\ (Paris) Lett.\ {\bf 38L},
355 (1977).
\bibitem{larson:01} R.~G.\ Larson, Macromolecules {\bf 34}, 4556 (2001).
\bibitem{milner:97} S.~T.\ Milner and T.~C.~B.\ McLeish,
Macromolecules, {\bf 30}, 2159 (1997).
\bibitem{likhtman:02} A.~E.\ Likhtman and T.~C.~B.\ McLeish,
 Macromolecules, {\bf 35}, 6332, (2002).
\bibitem{park:05a} 
S.~J.\ Park, S.\ Shanbhag and R.~G.\  Larson,
Rheol.\ Acta, {\bf 44},  319 (2005).

\bibitem{park:05b}  S.~J.\ Park and R~G Larson,
J. Rheol., {\bf 49}, 523 (2005).

\bibitem{larson:03} R.~G.\ Larson, T.\ Sridhar, L.~G.\ Leal, 
G.~H.\ McKinley, A.~E.\ Likhtman, and T.~C.~B.\ McLeish,
J. Rheol., {\bf 47}, 809 (2003).
\bibitem{fetters:pv} L.~J.\ Fetters, private communication.
\bibitem{read:01}D.~J.\ Read, and T.~C.~B.\ McLeish, Macromolecules {\bf 34},
1928 (2001).
\end{thebibliography}
\end{document}